%% file: main.tex
\documentclass[11pt]{article}
\usepackage{geometry}
\usepackage{setspace}
\usepackage{amsthm}
\usepackage{morefloats}

\usepackage{times}
\usepackage{graphicx,verbatim,array,multicol,fontenc,subfig,comment} 
\usepackage{psfrag, fancybox,color,soul} 
\usepackage{amsmath, amssymb, epsfig,multirow,bigstrut}
\usepackage{soul} \input{GrandMacros}

\usepackage{rotating}
\usepackage[table]{xcolor}
\definecolor{lightgray}{gray}{0.95}
\definecolor{ltgray}{gray}{0.9}
\usepackage{ dsfont }

\usepackage{natbib}

\newcommand{\noteC}[1]{{\footnote{\color{red} \bf{from Cory:} #1} }}

\newcommand{\appropto}{\mathrel{\vcenter{
  \offinterlineskip\halign{\hfil$##$\cr
    \propto\cr\noalign{\kern2pt}\sim\cr\noalign{\kern-2pt}}}}}

\title{Posterior Predictive Treatment Assignment for Estimating Causal Effects with Limited Overlap}

\author{Corwin M. Zigler$^{1}$ and Matthew Cefalu$^{2}$ \\ {\itshape $^{1}$Department of Biostatistics, Harvard TH Chan School of Public Health} \\{\itshape $^{2}$RAND Corporation}} 

\begin{document} 

\maketitle
\begin{abstract}
Estimating causal effects with propensity scores relies upon the availability of treated and untreated units observed at each value of the estimated propensity score.  In settings with strong confounding, limited so-called ``overlap'' in propensity score distributions can undermine the empirical basis for estimating causal effects and yield erratic finite-sample performance of existing estimators.  We propose a Bayesian procedure designed to estimate causal effects in settings where there is limited overlap in propensity score distributions.  Our method relies on the {\it posterior predictive treatment assignment} (PPTA), a quantity that is derived from the propensity score but serves different role in estimation of causal effects. We use the PPTA to estimate causal effects by marginalizing over the uncertainty in whether each observation is a member of an unknown subset for which treatment assignment can be assumed unconfounded. The resulting posterior distribution depends on the empirical basis for estimating a causal effect for each observation and has commonalities with recently-proposed ``overlap weights'' of \cite{li_balancing_2016}.  We show that the PPTA approach can be construed as a stochastic version of existing ad-hoc approaches such as pruning based on the propensity score or truncation of inverse probability of treatment weights, and highlight several practical advantages including uncertainty quantification and improved finite-sample performance.  We illustrate the method in an evaluation of the effectiveness of technologies for reducing harmful pollution emissions from power plants in the United States.

\end{abstract}


\section{Introduction}


Propensity score methods have gained widespread use for estimating causal effects with observational data.  Assuming that treatment assignment is ignorable (i.e., no unmeasured confounding), estimation of causal effects with the propensity score typically consists of two steps undertaken in sequence.  First, a model estimating the probability of receiving a particular treatment, conditional on covariates, is estimated. Predicted probabilities from this model, referred to as estimated propensity scores, are used to approximate the design of a randomized study where treatment is assumed to have been ignorably assigned within units having similar values of the estimated propensity score \citep{rosenbaum_central_1983, rubin_for_2008}.  Conditional on this ``design stage,'' causal estimates are obtained in an ``analysis stage'' by comparing outcomes between treated and untreated units with similar values of the propensity score.  Common implementations of these two stages include matching, stratifying, and inverse probability of treatment weighting (IPTW) by the propensity score to estimate causal effects \citep{rubin_for_2008, stuart_matching_2010, robins_marginal_2000}.  

A practical challenge to implementing propensity score methods arises in settings where strong confounding produces propensity score distributions that are very different in the treated and untreated units.  Such settings have been referred to as settings with limited ``overlap'' of propensity score distributions \citep{rubin_for_2008, crump_dealing_2009}.  We use the term ``limited overlap'' to describe settings where there is scarcity of observed treated or untreated units for some values of the propensity score, usually towards the tails of the propensity score distribution.  In extreme cases of limited overlap, treated or untreated units are unavailable for certain values of the propensity score.  Such a setting prevents causal inference without model-based extrapolation, and is sometimes referred to as a violation of the positivity assumption or experimental treatment assignment \citep{moore_ambient_2010}. Even when propensity score distributions do overlap, limited availability of treated units that can be compared against comparable (in terms of the estimated propensity score) untreated units can undermine practical usefulness of many common estimators.  The extent overlap relates to the empirical basis for causal inference, as values of the propensity score with many treated and untreated units contain more information than values with relatively few treated or untreated observations.  

Often, in the presence of complete lack of overlap, causal effects are estimated in a subset of the observed data by ``pruning'' or ``preprocessing'' the sample to omit observations with non-overlapping propensity scores \citep{ho_matching_2007}.  For settings with limited overlap, \cite{crump_dealing_2009} propose a method that, under certain assumptions, specifies a range of propensity score values determining the subset of units that will yield a minimum variance causal estimate. The problem of limited overlap has received ample attention  in the context of IPTW estimators, which remain asymptotically consistent but can be of limited practical value when units with limited overlap ``dominate the weighted analysis, with the result that [the] IPTW estimator will have large variance....reflect[ing] a lack of information in the data...'' (\cite{robins_marginal_2000}). Weight stabilization \citep{robins_marginal_2000} can alleviate this problem in some circumstances, but other approaches to truncate weights to a pre-specified maximum value or delete observations altogether have also been proposed to stabilize estimates \citep{kish_weighting_1992, cole_constructing_2008, moore_ambient_2010,kilpatrick_exploring_2013, xiao_comparison_2013}.  \cite{li_balancing_2016} offer an approach that does not truncate weights, but rather defines so-called ``overlap weights'' that will similarly avoid dominance of areas of the covariate distribution exhibiting limited overlap while focusing inference on the subset of units for which there is most empirical basis for making causal inference.

An important commonality among existing approaches to diagnose and mitigate the consequences of limited propensity score overlap is that these approaches ignore two sources of uncertainty.  First, judgments of overlap are based on the estimated propensity score as though this were a fixed an known quantity. Second, most decision rules to mitigate the consequences of limited overlap (e.g., through pruning or weight truncation) will necessarily involve the judgment of the analyst to adopt one of possibly many reasonable decisions.   For example, in a setting with strong confounding, a researcher may first estimate propensity scores and IPTWs, treat these values as fixed, and subjectively decide on a truncation point beyond which weights are deemed ``extreme.''  In this case, the truncation rule itself is one of many possible decisions, and applying this rule based on the estimated propensity score can conflate genuine confounding with pure estimation uncertainty in the propensity score; high variance in propensity score estimates can exacerbate limited overlap.

The objective of this paper is to develop methods to improve uncertainty quantification, finite-sample performance, and interpretability of causal estimators in the presence of limited overlap.  A Bayesian procedure is proposed that relies on the specification of a probability model for inclusion in the hypothetical ``design'' stage of the propensity score analysis.  This probability is linked to estimates of the propensity score and notions of overlap.  Conditional on the design stage, an analysis stage conducts a treatment contrast that is not confounded by observed covariates, reflecting estimates of a causal effect.  Ultimately, membership in the hypothetical study design is regarded as unknown for each observation, and the posterior distribution of the causal effect is estimated by marginalizing over the uncertainty in whether each observation is a member.  In accordance with recommendations such as those in \cite{rubin_for_2008}, the procedure separates ``design'' from ``analysis'' in the sense that estimation of each unit's membership in the design does not incorporate any information on the outcome.  The result is a posterior distribution of the causal effect where the contribution of each observation depends on the the empirical basis for estimating a causal effect for that observation. In this regard, the proposed approach can be construed as a stochastic version of existing ad-hoc approaches, such as weight truncation or deletion, pruning, or pre-processing: rather than rely on a single single arbitrary decision, the approach stochastically filters observations in a manner consistent with the principles underlying existing approaches while also acknowledging propensity score uncertainty.  This is particularly important, as changing the causal estimand of interest on the basis of arbitrary decisions of mathematical convenience (e.g., to stabilize estimates or yield a minimum variance estimate) can complicate scientific interpretation of the ultimate causal inference, in particular when such decisions are applied to an estimated quantity.  Methods that acknowledge the uncertainty associated with design decisions that change the causal estimand are, to our knowledge, nonexistent.  Furthermore, the proposed approach can be construed as a Bayesian  implementation of the overlap weights of \cite{li_balancing_2016} that acknowledges uncertainty in the estimated weights.

The proposed methods for estimating causal effects falls within a broader class of recent work aiming to incorporate the propensity score into Bayesian estimation procedures \citep{hoshino_bayesian_2008, mccandless_bayesian_2009, kaplan_two-step_2012, mccandless_adjustment_2012, zigler_model_2013, zigler_uncertainty_2014, zigler_central_2016}.  Section \ref{sec:desig_analysis} introduces notation and outlines the sequential design and analysis of observational studies with propensity scores.  Section \ref{sec:bayes} introduces the methodology for Bayesian stochastic pruning and marginalizing over this uncertainty in effect estimation.  Section \ref{sec:interpretation} addresses the interpretation of the proposed method and provides conceptual comparison with existing approaches.  Section \ref{sec:sims} evaluates the approach in a simple simulation study, and Section \ref{sec:application} applies the method to compare the effectiveness of power plant emission-control technologies for reducing harmful nitrogen oxide (NO$_x$) pollution.  We conclude with a discussion.

\section{``Designing'' and ``Analyzing'' Observational Studies with Propensity Scores} \label{sec:desig_analysis}
\subsection{Potential Outcomes, Ignorability, and Propensity Scores}\label{sec:framework}
Consider a dichotomous treatment $A=\{0,1\}$, a vector of covariates $\bX$, and potential outcomes $(Y_0,Y_1)$, where $Y_a$ denotes the outcome that would potentially be observed under treatment $A=a$. Define such quantities for each of $i=1,2,\ldots,n$ units, with $y_i$, $\bx_i$, and $a_i$ being the observed value of the outcome, covariate vector, and the treatment for the $i^{th}$ unit. Under the assumption of strong ignorability (i.e. $(Y_0,Y_1) \bot A | \bX$), observed data can be used to estimate the causal effect of $A$ on $Y$ by comparing outcomes on treated and untreated units with similar values of $\bX$. 

The propensity score is often employed as a tool for condensing the relevant information in $\bX$ into a scalar-valued \textit{balancing score} such that, under the assumption of strong ignorability, average outcomes in treated and untreated units with similar values of the propensity score represent estimates of causal effects \citep{rosenbaum_central_1983}. A variety of implementations are available, all with the salient theme that conditioning on the estimated propensity score renders the treatment assignment unconfounded, permitting conditional comparisons between observed outcomes on treated and untreated units to be interpreted as causal contrasts between common sets of units.

Procedures based on the propensity score can target various causal estimands such as the population average treatment effect (ATE) or the average treatment effect on the treated (ATT).  In some instances, initial interest lies in a specific estimand such as the ATE, but interest shifts to alternative causal quantities when limited propensity score overlap undermines estimation of the ATE. In other circumstances, causal quantities other than the ATE are of intrinsic interest.

\subsection{The ``Design'' Stage: Unconfounded Subsets of the Observed Data}
Propensity score approaches are frequently framed as first using the propensity score to approximate the hypothetical design of a randomized experiment then, conditional on this ``design stage,'' evaluating causal effects.  Many such decisions may be involved in the design stage, including which variables to include in propensity score estimation and how to construct matches or subgroups of comparable units.  When confronted with extreme confounding and limited propensity score overlap, one design decision is whether exclude or otherwise alter the contribution of units exhibiting limited overlap. A successful design stage renders the treatment assignment unconfounded conditional on the design decisions.  

For any set of $n$ observational units, consider the existence of a subset of the observations for which treatment assignment could be regarded as unconfounded, that is, a subset of observations that would approximate the design of an experiment that would permit estimation of a causal effect of $A$ on $Y$ (i.e. a subset within which strong ignorability holds).  Let the indicator $S_i=\{0,1\}$ denote whether the $i^{th}$ observation is included as a member of such an unconfounded subset, and let $\Ssc \equiv \{i: S_i=1\}$ represent the entire subset. If $A$ were randomly assigned as in a controlled experiment, then $\Ssc$ would be known to consist of the entire sample, with $S_i = 1$ for all $i = 1, 2, \ldots, n$.  When treatment is not randomized, as in an observational study, $\Ssc$ can be regarded as an unknown subset of the observed sample.  We describe our procedure for marginalizing inference for causal effects over the uncertainty in $\Ssc$ in Section \ref{sec:bayes}, but note here that such an $\Ssc$ is implied by many existing procedures for estimating causal effects with propensity scores.   Omitting observations (via pruning or weight deletion) that are deemed to have non-overlapping or close to non-overlapping propensity scores corresponds to setting $S_i=0$ for these observations.

Note that the design stage described above is intentionally simplistic so that it can serve as a device for framing the proposed methodology focusing specifically on mitigating the consequences of limited propensity score overlap, and is not meant to represent a comprehensive description of the wide variety of design decisions that may be employed to approximate a hypothetical experiment with observational data.  



\subsection{The ``Analysis'' Stage: Models for Observed Data}\label{sec:analysis_stage}
If $\Ssc$ were known, a causal contrast could be defined, conditional on $\Ssc$, with any observed-data contrast between treated and untreated observations. Let $\Delta$ refer generically to the causal estimand defined by this contrast.  While $\Delta$ could be specified in many ways, one common and useful way is to specify a marginal structural model for potential outcomes of the form:
\begin{align}
     \E(Y_a) = \theta + \Delta a \label{eqn:msm},
\end{align}
where the causal estimand of interest is average treatment effect (ATE), $\Delta$.  We focus on this specification of a causal contrast for illustration throughout, but the key idea is that, conditional on $\Ssc$ where treatment is unconfounded, observed data on treated and untreated units would permit estimation of $\Delta$ with a model analogous to (\ref{eqn:msm}) but fit to observed data:
\begin{align}
     E(Y|A=a,S=1) = \beta_0 + \beta_1 a \label{eqn:msm_obs},
\end{align}
where $\beta_1$ shares the same interpretation as $\Delta$.

\section{Bayesian Estimation: Integrating Design Uncertainty into Analysis}\label{sec:bayes}
\subsection{General formulation of marginalizing over design uncertainty}
If $\Ssc$ were known (e.g., in a controlled experiment) or fixed (e.g., through a process for pruning), Bayesian inference for the causal effect of $A$ on $Y$ could follow from a likelihood function of the form $\Lsc(\by | \ba , \bx , \Ssc, \beta)$ consistent with the model in Equation (\ref{eqn:msm_obs}).  Note that the conditioning on $\bx$ indicates that the model may add adjustment of individual covariates, $\bx$, but we forego such adjustment for simplicity. It is important to note that the role of $\Ssc$ in $\Lsc(\by | \ba , \bx , \Ssc, \beta)$ is akin to that of a sample-selection indicator, and conditioning on $\Ssc$ is assumed to satisfy strong ignorability and permit observed-data causal contrasts with the specified likelihood (with or without direct adjustment for $\bx$).   Combining this likelihood with a prior distribution for $\beta$, $\pi(\beta)$, posterior inference for the causal effect would follow from  $p(\beta | \by, \ba ,  \bx , \Ssc) \propto  \Lsc(\by | \ba , \bx , \Ssc, \beta) \pi(\beta)$.

In observational studies, $\Ssc$ is typically unknown.  Existing approaches that condition all inference for causal effects on the design stage can be regarded as fixing $\Ssc$ \textit{a priori}.  In principle, Bayesian inference can acknowledge the intrinsic uncertainty surrounding $\Ssc$ through integrating over an unknown probability distribution for $\Ssc$:
 \begin{align} 
	p(\beta | \by, \ba ,\bx)  & \propto \int \Lsc(\by | \ba , \bx , \Ssc, \beta) p(\Ssc |\ba,\bx ) \pi(\beta) \partial \Ssc \label{eqn:marg},
\end{align}
where, in the present context, the model for $\Ssc$ depends only on quantities in the design stage. Note that, for simplicity of exposition, Equation (\ref{eqn:marg}) assumes that the parameter $\beta$ is the same for all values of $\Ssc$, which would also render $\beta_1 = \Delta$.  This will not hold in general, which will be discussed in detail in Section \ref{sec:interpretation}. 

As written, there is no inherent difficulty to evaluating a posterior distribution of the form of that in (\ref{eqn:marg}).  Other problems can be cast in a similar form, for example, if $\Ssc$ were simply regarded here as a missing or mismeasured covariate, and $p(\Ssc|\ba, \bx)$ a model for its distribution.  However, as will be elaborated, anchoring $\Ssc$ to the propensity score, which is itself unknown and must be estimated, introduces fundamental and unique complications to the evaluation of a posterior such as (\ref{eqn:marg}).

\subsection{A model for the propensity score}\label{sec:ppps}
Let $\prod_{i=1}^n\Lsc(a_i | \bx_i, \bgamma)$ denote a likelihood function for the mechanism governing treatment assignment, that is, for a propensity score model indexed by an unknown parameter $\bgamma$ having prior distribution $\pi(\bgamma)$.  As noted in \citet{rosenbaum_central_1983}, estimated propensity scores could be regarded in a Bayesian paradigm as posterior-predictive probabilities of treatment assignment derived from $p(\bgamma|\ba, \bx) \propto \prod_{i=1}^n\Lsc(a_i | \bx_i, \bgamma)\pi(\bgamma)$, but only recently have methods begun to consider how to incorporate the posterior-predictive distribution of the propensity score into inference for causal effects \citep{hoshino_bayesian_2008, mccandless_bayesian_2009, kaplan_two-step_2012, zigler_model_2013, zigler_uncertainty_2014, zigler_central_2016, alvarez_uncertain_2014,spertus_bayesian_2017}.  




\subsection{A model for $\Ssc$ using the posterior-predictive treatment assignment (PPTA)}\label{sec:ppta}
Since the propensity score encodes information about the probability of receiving the treatment for a unit with $\bX = \bx$, it captures in part the availability of units with both values of $A$ for that value of $\bx$; values of $\bx$ leading to propensity estimates close to 1 suggest the scarcity of untreated units with those characteristics. In this sense, the probability that a unit could have received the {\it opposite} treatment characterizes the extent of overlap at that point in the propensity score distribution, and relates directly to the empirical basis for making a causal inference for that unit.  This motivates anchoring a probability model for $\Ssc$ to estimates of the propensity score to ensure that inclusion in the hypothetical study design relates to the number of both treated and untreated observations at a given value of the propensity score. 

Rather than consider the posterior-predictive propensity score directly, define the posterior-predictive treatment assignment (PPTA) as a realization from a Bernoulli random variable, with probability of success equal to the posterior-predictive propensity score for each observation. That is, denote the PPTA for observation $i$ with $\Atilde_i \sim Bernoulli\left(p=P(A=1|\bX=\bx_i, \gamma)\right)$, where $P(A=1|\bX=\bx_i, \gamma)$ the corresponding posterior-predictive probability for a unit with $\bX=\bx_i$ and a particular value of $\gamma$ (which will be simulated from its posterior distribution).  One probability model for $\Ssc$ (conditional on a realization of the posterior-predictive propensity score) can be specified as:

\begin{align}
p(\Ssc | \bx, \ba, \gamma) = \prod_{i=1}^n p(S_i=1 | \bX=\bx_i, A=a_i, \gamma) &= \prod_{i=1}^n p(\Atilde_i = 1-a_i | \bX=\bx_i, A=a_i, \gamma) \label{eqn:smod1}.
\end{align}
Anchoring the probability distribution for $\Ssc$ to the probability that $\Atilde_i \ne a_i$ relates to the extent of overlap and the empirical basis for causal inference.  Treated observations with high propensity scores would have few untreated observations with similar propensity score values, and, correspondingly, a low frequency of $\Atilde_i \ne a_i$, and a low probability of $S_i = 1$.  In contrast, observations with low (high) propensity scores that happened to be treated (untreated) can be regarded as very likely to have received the opposite treatment, which would manifest as frequent simulations of $\Atilde_i \ne a_i$.  Observations with propensity scores closer to 0.5 are likely to have received either treatment, and will also have substantial posterior probability that $\Atilde_i \ne a_i$.  Thus, observations with a high probability of inclusion in $\Ssc$ are those with values of $\bx$ that suggest a reasonably high likelihood of having received the treatment opposite of that observed which, in practice, represents observations that are similar to many observations in the opposite treatment group. 


Prioritizing inclusion of observations in $\Ssc$ with (\ref{eqn:smod1}) is akin to ``smoothing'' the hypothetical study design towards one with equal probability of receiving treatment for every value of the propensity score, which is similar to the small-sample exact balancing property of overlap weights noted in \cite{li_balancing_2016}.

\subsection{Two-Stage Bayesian Effect Estimation with the PPTA}\label{sec:twostage}
While the discussion in Section \ref{sec:ppta} focuses on models for $\Ssc$ using a single realization from the posterior-predictive propensity score distribution, the ultimate goal is to integrate uncertainty in the propensity score and in $\Ssc$ into inference for causal effects in the analysis stage, as represented heuristically in expression (\ref{eqn:marg}).  The key challenge is that defining $\Ssc$ based on the estimated propensity score cannot naturally be expressed in the context of a joint distribution for $p(\by, \bS, \bx, \ba, \beta, \gamma)$, which is required in a standard Bayesian analysis that factors the joint probability distribution of all observable quantities into a likelihood conditional on unknown parameters and a prior distribution for these parameters \citep{zigler_central_2016, zigler_model_2013, robins_toward_1997, robins_discussion_2015}. This motivates a two-stage procedure for estimating the marginal posterior of $\beta$ in (\ref{eqn:marg}) that maintains the distinction between design and analysis stages while allowing uncertainty in the former to propagate into inference in the latter.  

The first estimation stage corresponds to the ``design'' stage.  Bayesian estimation of the parameters in the propensity score model and $\Ssc$ characterize each unit's inclusion in the hypothetical study design. This is done using only information pertinent to the design, i.e., without use of the outcome. The second estimation stage corresponds to the ``analysis'' of the hypothetical study designed in the first stage.  For a given realization of $\Ssc$ in the first stage, an unweighted and unadjusted analysis is conducted as outlined in Section \ref{sec:analysis_stage} using $\Lsc(\by | \ba , \bx , \Ssc, \beta)=\prod_{i} \Lsc(y_i | a_i , \bx_i , \beta)^{S_{i}}$. This outcome likelihood in stage two is iteratively evaluated for posterior simulations of $\Ssc$ from stage one, thus marginalizing over uncertainty in the design stage when estimating effects in the analysis stage.  

More formally, such a procedure corresponds to the following expression of the posterior distribution of the causal estimand:

\begin{align} 
	p(\beta | \by, \ba, \bx) &\propto  \int _\Ssc  \underbrace{p(\beta |\by, \ba,  \bx , \Ssc)}_{\mathrm{``Analysis''}} \underbrace{p(\Ssc|\bx, \ba)}_{\mathrm{``Design''}} \partial \Ssc \notag \\
	&= \int_\Ssc \underbrace{ \pi(\beta) \prod_{i=1}^n  \Big\{ \Lsc(y_i | a_i , \bx_i , \beta)^{S_{i}}}_{\mathrm{``Analysis''}} \underbrace{p(S_i=1 | \bx_i, a_i)  }_{\mathrm{``Design''}}  \Big\} \partial \Ssc \notag \\
	&= \int_\Ssc \underbrace{ \pi(\beta) \prod_{i=1}^n  \Big\{ \Lsc(y_i | a_i , \bx_i , \beta)^{S_{i}}}_{\mathrm{``Analysis''}} \underbrace{\Big[ \int_{\gamma} p(S_i=1 | \bx_i, a_i, \gamma)p(\gamma|\bx, \ba) \partial \gamma\Big] }_{\mathrm{``Design''}}\Big\}  \partial \Ssc \notag \\
	&= \int_\Ssc \underbrace{ \pi(\beta) \prod_{i=1}^n  \Big\{ \Lsc(y_i | a_i , \bx_i , \beta)^{S_{i}}}_{\mathrm{``Analysis''}} \underbrace{ \Big[ \int_{\gamma} p(S_i=1 | \bx_i, a_i, \gamma) \left( \prod_{j=1}^n \left[ \Lsc(a_j|\bx_j, \gamma) \right] \pi(\gamma) \right) \partial \gamma \Big]}_{\mathrm{``Design''}} \Big\} \partial \Ssc \label{eqn:fullposterior} 
\end{align} 

Looking at the last line in (\ref{eqn:fullposterior}), the term in the large parentheses with the likelihood product over $j$ is simply the posterior distribution of the parameters of the propensity score model, evaluated without regard to quantities in the analysis stage.  Simulations from this posterior distribution (represented with the integration with respect to $\gamma$) generate the posterior-predictive propensity score distribution.  Posterior-predictive propensity scores are used to generate the probabilities $p(S_i=1|\bx_i, a_i, \gamma)$, specified as in Section \ref{sec:ppta}.  This constitutes the design stage that produces posterior simulations of $\Ssc$. 

The part of the likelihood corresponding to the analysis stage (product over $i$) is evaluated conditional on $S_i$ to estimate the posterior distribution of $\beta$.  Note that this portion of the likelihood is agnostic with respect to specific values $\gamma$ (and the estimated propensity scores), which is marginalized over in the portion of (\ref{eqn:fullposterior}) corresponding to the design stage. Only the simulations of $S_i$ are passed from the design stage to analysis stage, with these simulations marginalized over all uncertainty in $\gamma$ (and the propensity scores). Confining the marginalization over $\gamma$ to the design stage propagates uncertainty in the propensity score and in $\Ssc$ into the analysis stage.  However, note that the posterior distribution of $\gamma$ (and hence $\Ssc$) is not updated based on the data and parameters in the analysis stage, implying that the inference in the analysis stage does not contribute to the construction of the hypothetical study design \citep{rubin_for_2008}.  Such two-stage estimation of causal effects with propensity scores, without explicit consideration of $\Ssc$, has been previously considered previously \citep{mccandless_cutting_2010, zigler_model_2013, zigler_central_2016}, often framed as the need to avoid ``feedback'' between the design and analysis stage.  This is also a special case of modularization in Bayesian inference \citep{liu_modularization_2009, jacob_better_2017}.

\subsection{Computational Procedure}\label{sec:computation}
In practice, estimation of the posterior distribution in (\ref{eqn:fullposterior}) can be achieved with a simple Markov chain Monte Carlo (MCMC) algorithm carried out in two steps in accordance with the separation of the design and analysis stages. First, the posterior distribution of the parameters in the propensity score model ($p(\gamma|\bx, \ba)$) is obtained with standard approaches to construct an MCMC chain of length $m_1$. Simulations from this posterior can be used to generate $m_1$ posterior simulations of $\Ssc$, denoted with $(\Ssc_1, \Ssc_2, \ldots, \Ssc_{m_1})$.  For each simulation, $\Ssc_m$, a separate MCMC chain of length $m_2$ is run to simulate from the conditional posterior $p(\beta|\bx, \by, \ba, \Ssc_m)$, using only observations with $S_i=1$.  Simulations from $p(\beta|\bx, \by, \ba, \Ssc_m)$ are combined for all $m_1$ simulations of $\Ssc$ to obtain the marginal posterior $p(\beta|\bx, \by, \ba)$ using all $m_1 \times m_2$ simulations. Note that checking covariate balance can be conducted within the MCMC for each simulated value of $\Ssc_m$, which we illustrate in Section \ref{sec:application}.

\section{Interpretation and Comparison of PPTA with Existing Approaches}\label{sec:interpretation}

The ultimate result of estimating causal effects with (\ref{eqn:fullposterior}) is that each unit's contribution to the posterior distribution of $\beta$ differs depending on the posterior probability of that unit's inclusion in the hypothetical design stage.  Observations exhibiting the most empirical basis for causal inference (as determined by the extent of comparable observations with the opposite treatment) will be included most often in the hypothetical design and will contribute the most information to the final posterior inference.  Observations exhibiting limited overlap will contribute less to the inference for causal effects. 

Causal contrasts conditional on $\Ssc$ (even if $\Ssc$ were known) are not necessarily interpretable as the population ATE, owing to possible differences in the treatment effect between units in $\Ssc$ and those who are excluded. The election to forego estimation of the ATE in favor of estimating causal effects conditional on $\Ssc$ is one essential feature of the proposed approach and indeed all existing approaches designed to mitigate the consequences of limited propensity score overlap.  In exchange for changing the causal estimand, the proposed approach and other existing approaches aim to improve empirical reliability and practical usefulness by targeting an estimand that may be related to the effect of interest.  


More formally, the parameter $\beta_1$ in the analysis model (\ref{eqn:msm_obs}) is not guaranteed to represent the same estimand for different values of $\Ssc$, nor is it guaranteed interpretation as $\Delta$.  Thus, without the assumption that $\beta$ remains the same for all $\Ssc$, the representation in (\ref{eqn:marg}) can be related to the posterior distribution of $\Delta$ with:
\begin{align}
p(\Delta|\by,\ba,\bx) \appropto  \int \Lsc(\by | \ba , \bx , \Ssc, \beta_\Ssc) p(\Ssc |\ba,\bx ) \pi(\beta_\Ssc) \partial \Ssc \label{eqn:approxpost},
\end{align}
where the parameter denoting the causal contrast in the analysis stage is now acknowledged to explicitly depend on which observations are included in the design and, for simplicity, we abuse notation and omit distinction between $\beta_0$ and $\beta_1$. The extent of approximation in (\ref{eqn:approxpost}) relates to the nature of variability in the true underlying causal effect across different values of the propensity score.  If there is heterogeneity in the treatment effect across different values of $\Ssc$, relying on the approximation in (\ref{eqn:approxpost}) will provide biased estimates of the ATE, $\Delta$. 

Regardless of the extent of approximation of the ATE, the marginal posterior causal effect estimate in (\ref{eqn:approxpost}) with the proposed specifications of $p(\Ssc|\bx, \ba)$ is a causal contrast that may be of intrinsic interest due to its weighting of units based on the propensity to receive the {\it other} treatment.  \cite{li_balancing_2016} define such a quantity as the Average Treatment Effect for the Overlap population (ATO), which may be of interest for its pertinence to the population having characteristics that could have received either treatment with substantial probability. For example, in the context of the air quality intervention analyzed in Section \ref{sec:application}, interest may lie primarily in the power plants for which there is reasonable choice of whether to install an emissions control technology, eschewing focus on those in the population that will almost definitely (or definitely not) invest in a technology due to their size, operating efficiency, and/or usage pattern.

All approaches to mitigate the consequences of limited overlap, including the proposed approach, are met with the inherent feature that inference is redirected away the population ATE and towards an alternative quantity that can be estimated more reliably. The key feature of the proposed approach is that rather than confine interest to a single subset of the data of unclear interpretation (e.g., by pruning or weight truncation), inference is marginalized over many such subsets in a manner that renders interpretation as the ATO.  When the ATO is of particular relevance to the setting at hand, this induces no interpretational challenge.  When the ATE is the quantity of primary interest, any discordance between the ATE and the ATO amounts to a standard bias-variance tradeoff; the marginal posterior effect estimate, while possibly biased for the ATE, will be of the quantity that can be best estimated by the data at hand, possibly yielding finite-sample performance superior to other strategies that are asymptotically unbiased for the ATE.  Importantly, unlike existing approaches, the proposed approach also acknowledges the uncertainty in the estimated propensity scores themselves, meaning that not only is the rule for determining membership in the study design stochastic, the quantity to which that rule applied also varies with ordinary estimation uncertainty.

\section{Simulation study} \label{sec:sims}
To investigate the behavior of the proposed PPTA approach in comparison with alternative approaches to mitigate the consequences of limited overlap, we proposed a series of simulation studies that mimic differing levels of overlap in the covariate distributions between the treatment groups.  For the purposes of comparison, we assume a setting where the treatment effect is the same for all observations in the population, so that the ATE and the ATO share the same value and the relationship in (\ref{eqn:approxpost}) is not an approximation.  

\begin{figure}[htbp]
    \centering
    \includegraphics[width=\linewidth]{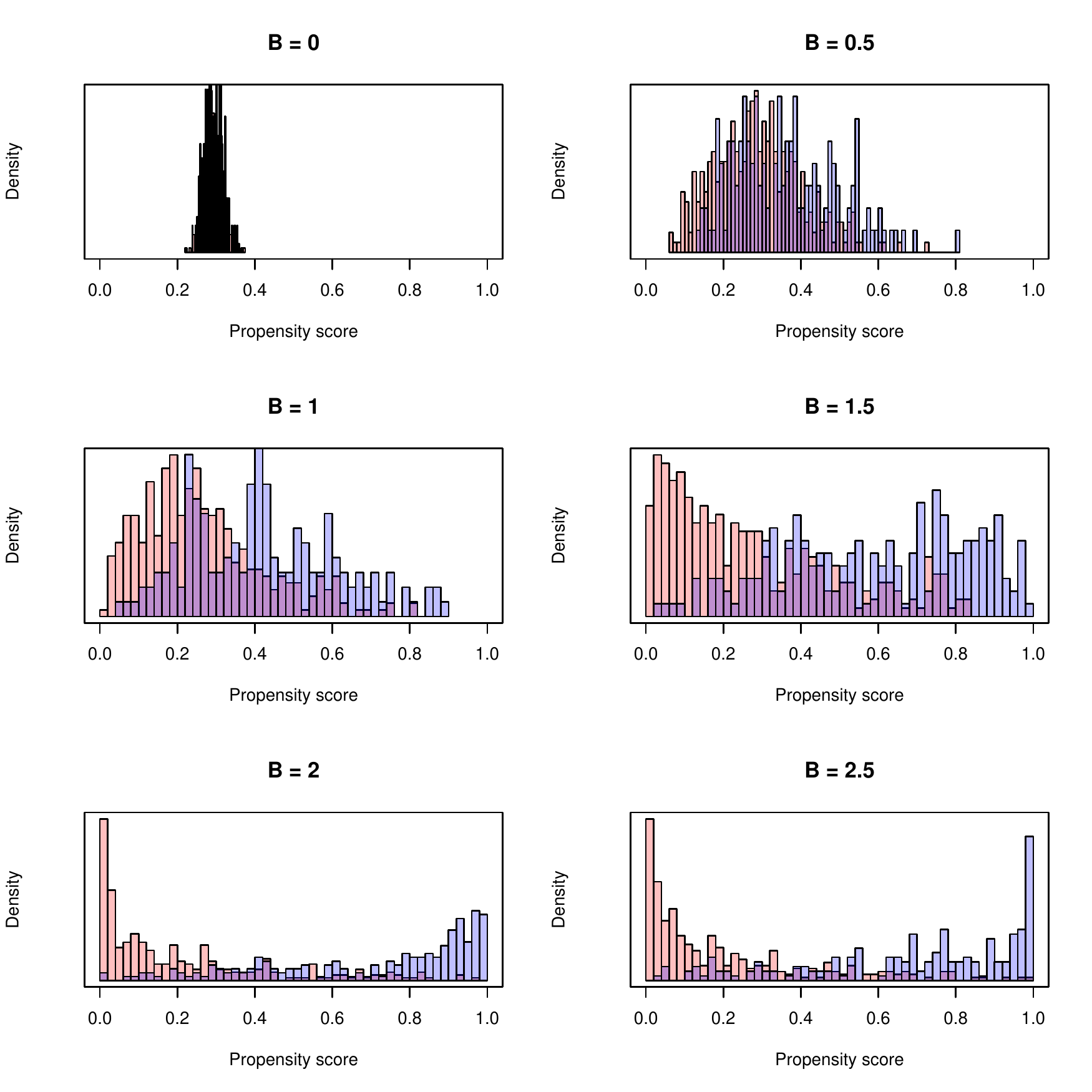}
    \caption{The propensity score distribution for each treatment group for varying choices of $B$.}
    \label{fig:PSdist}
 \end{figure}

Specifically, we generate 500 datasets in the following manner:

\begin{enumerate}
	\item For $i=1,...,500$, generate 5 independent, normally distributed covariates, $\bX_i$, with mean 0 and variance 1 
	\item Denote the first eigenvector of $\bX_i$ as $\bU_i$
	\item Generate the treatment under the following model: $A_i \sim Bernoulli( p_i=expit( \log(\frac{.3}{.7}) + BU_i ) )$
	\item Generate the outcome under a normal linear model: $Y_i \sim \Nsc( 0.2A_i + (-1)\times U_i , 1)$
\end{enumerate}

\noindent where $B=\{0,1/2,...,5/2\}$. As $B$ increases, confounding increases and extent of covariate overlap decreases. Figure \ref{fig:PSdist} displays the propensity score distributions for treated and untreated observations for one simulated data set under each value of $B$.  


We analyze each simulated data set with the PPTA approach described in Section \ref{sec:bayes}, implemented with an MCMC chain of length 1500 (discarding the first 1000 as burn in) for estimating the propensity score model and $\Ssc$, combined with an MCMC chain of length 1500 (discarding the first 1000 as burn in) for each simulated value of $\Ssc$, totaling $500 \times 500 = 250000$ MCMC iterations used to construct the posterior distribution of the ATE. We compare the proposed PPTA approach with IPTW with and without weight truncation, the approach of \cite{crump_dealing_2009}, and the overlap weights of \cite{li_balancing_2016} to estimate the ATO. Three different IPTW truncation values were considered (100, 50, 10), and a comparison with weight deletion is omitted due to its similar performance to truncation. 



\subsection{Simulation results}

\begin{figure}[htbp]
\caption{Simulation results: comparison of methods with normally-distributed outcome}\label{fig:gaussian}
\includegraphics[width = \textwidth]{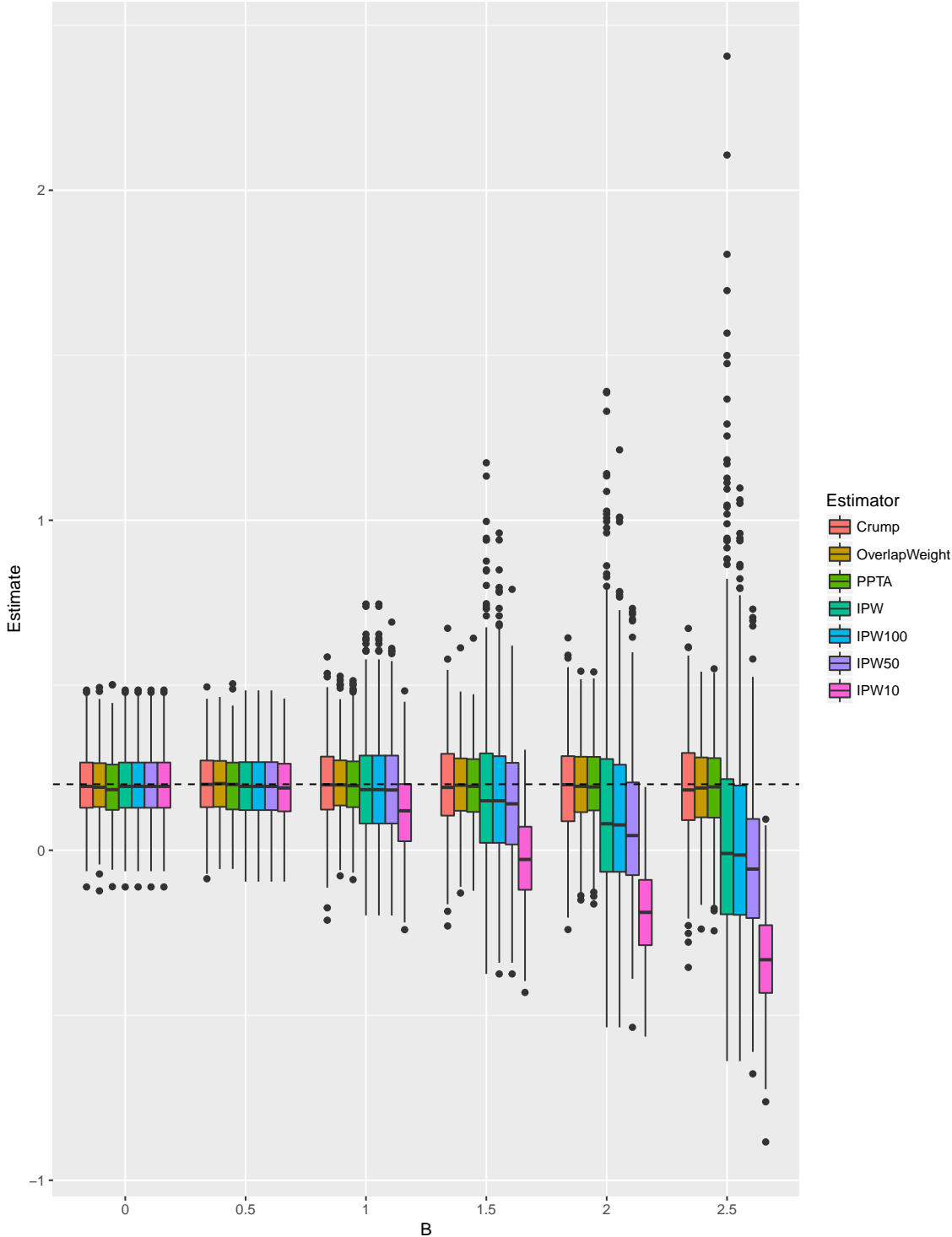}
\end{figure}

Figure \ref{fig:gaussian} provides boxplots of point estimates of the ATE using each approach described above, calculated across 500 simulated data sets datasets for each specified value of $B$ (point estimates are posterior means from PPTA). For $B=0$ or $0.5$, all estimators have similar distributions, suggesting that their overall performance is similar in data sets simulated to have little or no confounding.

For $B=1$ and $B=1.5$, performance of the IPTW approaches begin to suffer.  Truncating weights at 10 leads to bias in the point estimates, while larger truncation points (or no truncation) exhibit approximate unbiasedness but increased variability relative to the PPTA, overlap weights, and approach of \cite{crump_dealing_2009}.  For $B>1$, the erratic performance of IPTW with and without truncation becomes increasingly evident as $B$ increases, with even the IPTW estimator without truncation exhibiting finite-sample bias and extreme variability.  Different truncation values show less variability but incur bias as $B$ increases.  The PPTA, overlap weights, and \cite{crump_dealing_2009} estimates remain unbiased, with the PPTA and overlap weights approaches exhibiting the least amount of variability in point estimates.

\section{Evaluating the Effectiveness of Power Plant Emissions Controls} \label{sec:application}
Years of epidemiological research have linked ambient ozone pollution to adverse health outcomes \citep{bell_ozone_2004, jerrett_long-term_2009}.  As a consequence, a variety of regulatory strategies in the U.S. are designed to reduce ambient ozone pollution through incentivizing electricity-generating units (i.e., ``power plants'') to reduce emissions of nitric oxide and nitrogen dioxides (NO$_x$), important precursors to the formation of ambient ozone.  While many technologies for reducing NO$_x$ emissions are available, Selective Catalytic Reduction (SCR), and Selective Non-Catalytic Reduction (SNCR) technologies are believed to be among the most efficient for reducing NO$_x$.  However, empirical verification of the comparative effectiveness of SCR/SNCR systems for reducing NO$_x$ emissions (relative to alternatives) is limited. 

To illustrate the proposed PPTA approach, the  subsequent analysis estimates, for each year from 2002-2014, the effectiveness of SCR/SNCR technologies (relative to other strategies) for reducing annual NO$_x$ emissions.  The number of units considered per year ranges from 1505 (in 2003) to 2028 (in 2008).  Prevalence of SCR/SNCR technologies ranges from 15\% (in 2002) to 47\% (in 2013 and 2014). Each year is considered separately, as the effect of emission-control technologies on emissions is virtually instantaneous with effects not expected to vary from year-to-year or build up over time. For each year, we estimate the causal effect of SCR/SNCR on NO$_x$ emissions using five approaches: standard IPTW, IPTW with weights truncated at 50 (IPTW50), the method of \cite{crump_dealing_2009}, the overlap weights of \cite{li_balancing_2016}, and the proposed PPTA approach.

\subsection{Limited Overlap in the Evaluation of NO$_x$ Emissions-Reduction Technology}
A key challenge to evaluating SCR/SNCR technology with observed data is the strength of confounding.  Power plant characteristics such as size, operating capacity, and fuel type are strongly associated with whether a power plant elects to install SCR/SNCR technology.  What's more, a string of regulations starting with at least the Acid Rain Program in the early 2000s have continued to strengthen rules for NO$_x$ emissions, prompting increased SCR/SNCR technology use in later years.  While the effectiveness of the systems for reducing NO$_x$ emissions is not expected to change over time, the extent of confounding does: earlier years exhibit more confounding since the relatively weaker regulations prompted only certain types of plants to invest in SCR/SNCR technology (e.g., older, larger plants) while tighter regulations in later years prompted investment in such technology among a wider range of plants (once the technology is installed, it is not removed).  Thus, at any given year and especially in earlier years, estimation of the population ATE of SCR/SNCR technology for reducing NO$_x$ emissions is confronted with limited propensity score overlap, and a quantity such as the ATO is of particular interest for its focus on the types of plants that might plausibly adopt SCR/SNCR technology given the current regulatory environment.  

\begin{figure}[htbp]
\subfloat[2002]{
  \includegraphics[width=.5\textwidth]{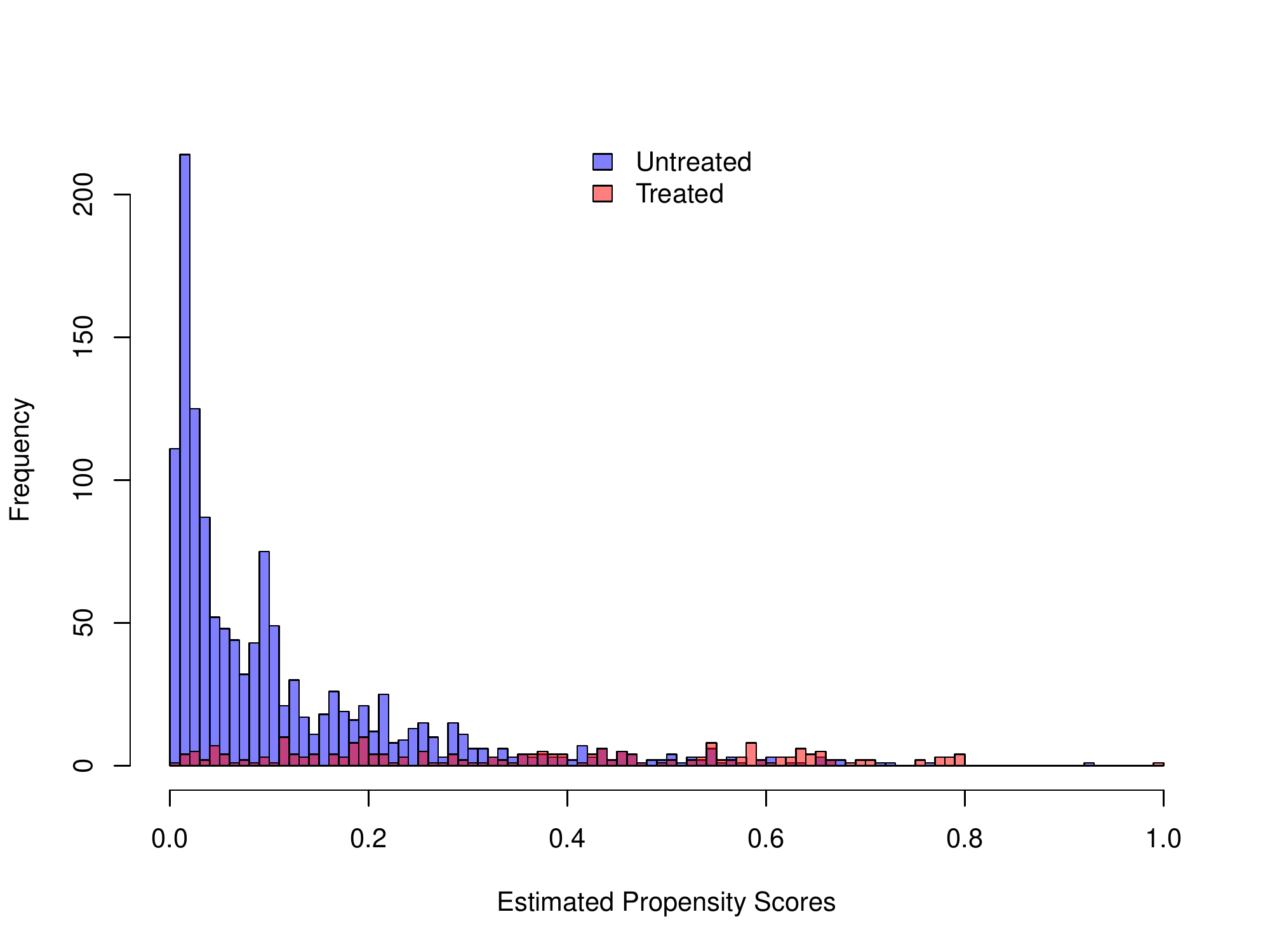}\label{fig:ps2002}}
\subfloat[2003]{
  \includegraphics[width=.5\textwidth]{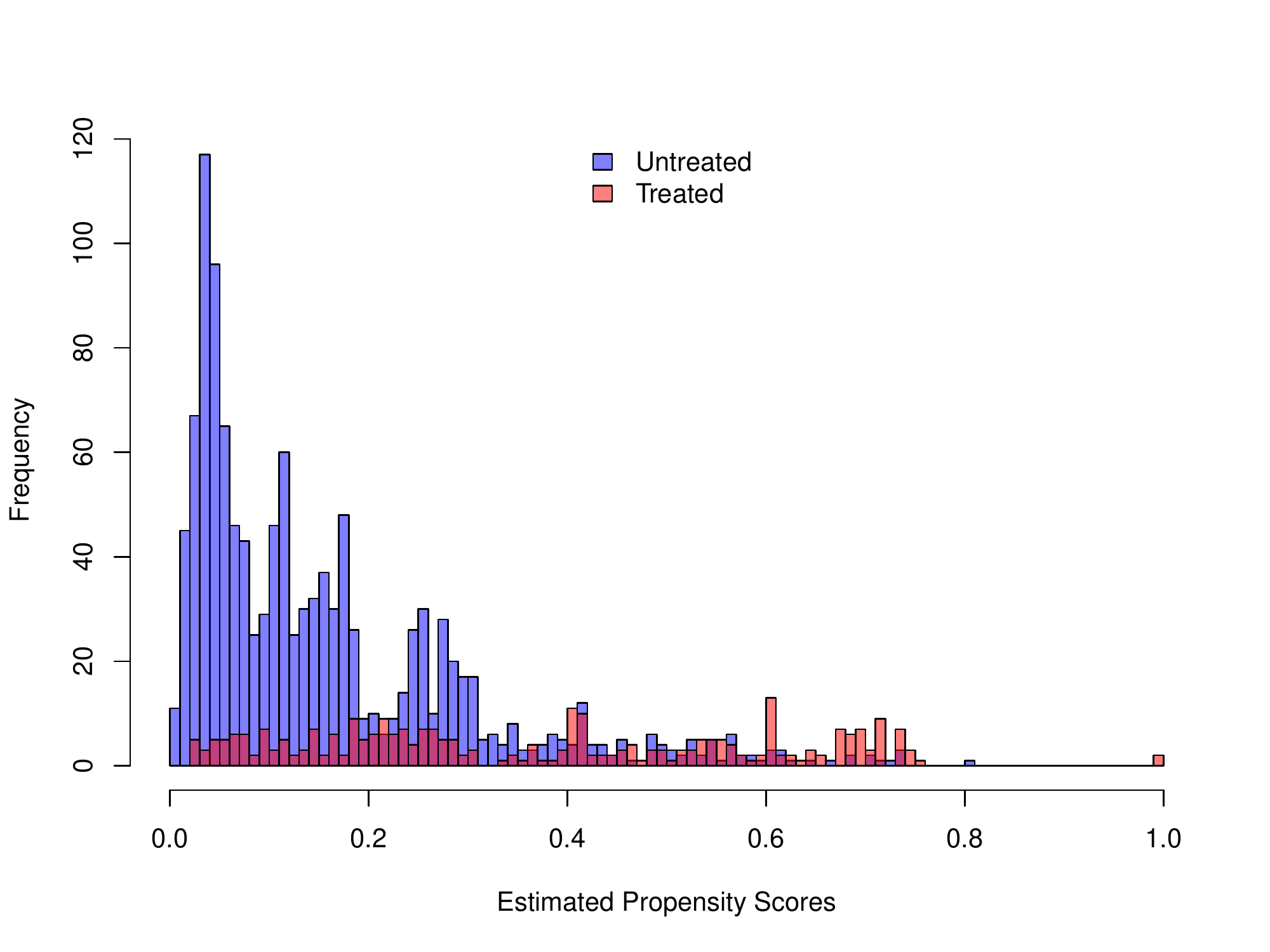}\label{fig:ps2003}}\\
\subfloat[2008]{
  \includegraphics[width=.5\textwidth]{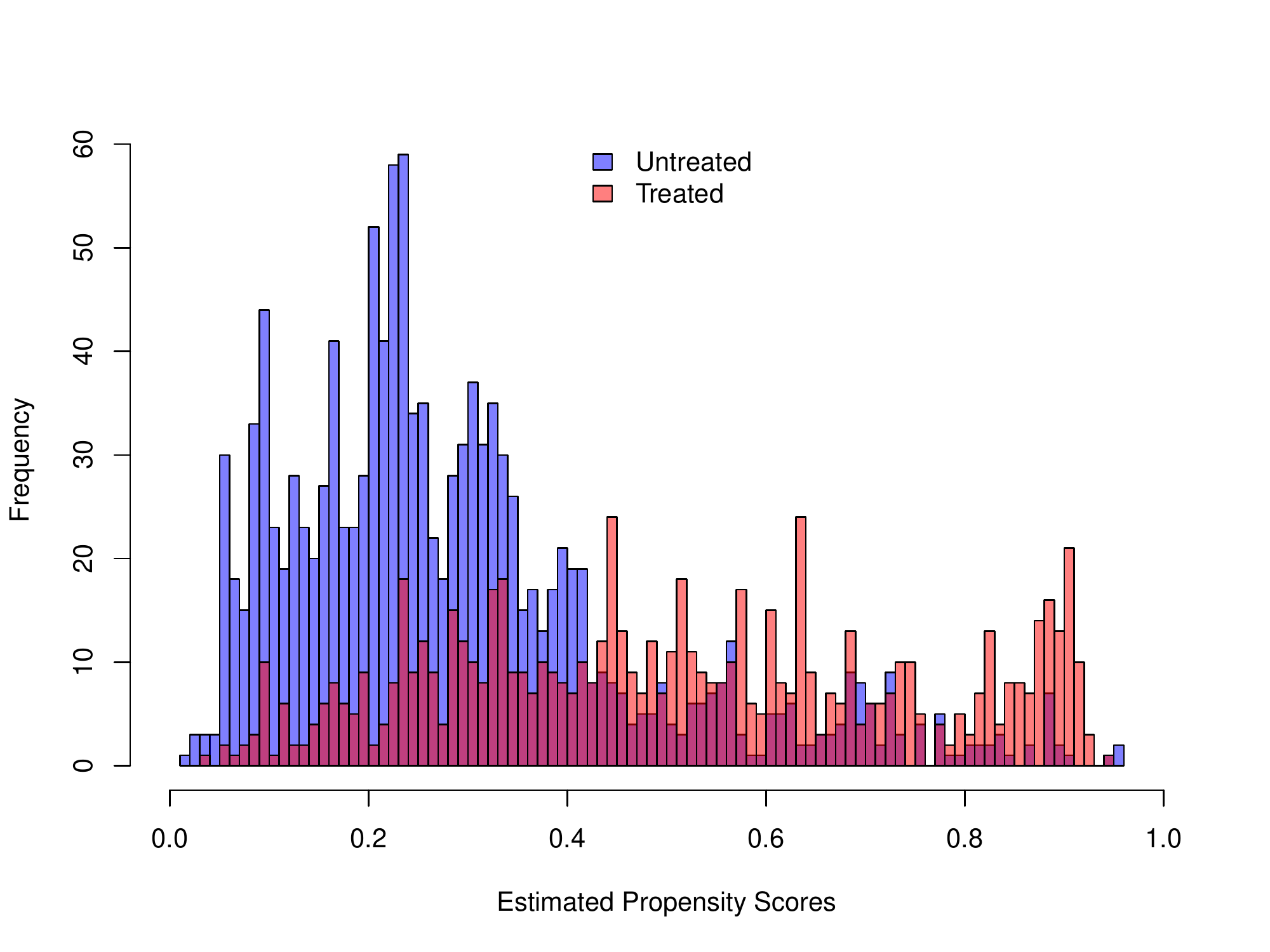}\label{fig:ps2008}}
\subfloat[2014]{
  \includegraphics[width=.5\textwidth]{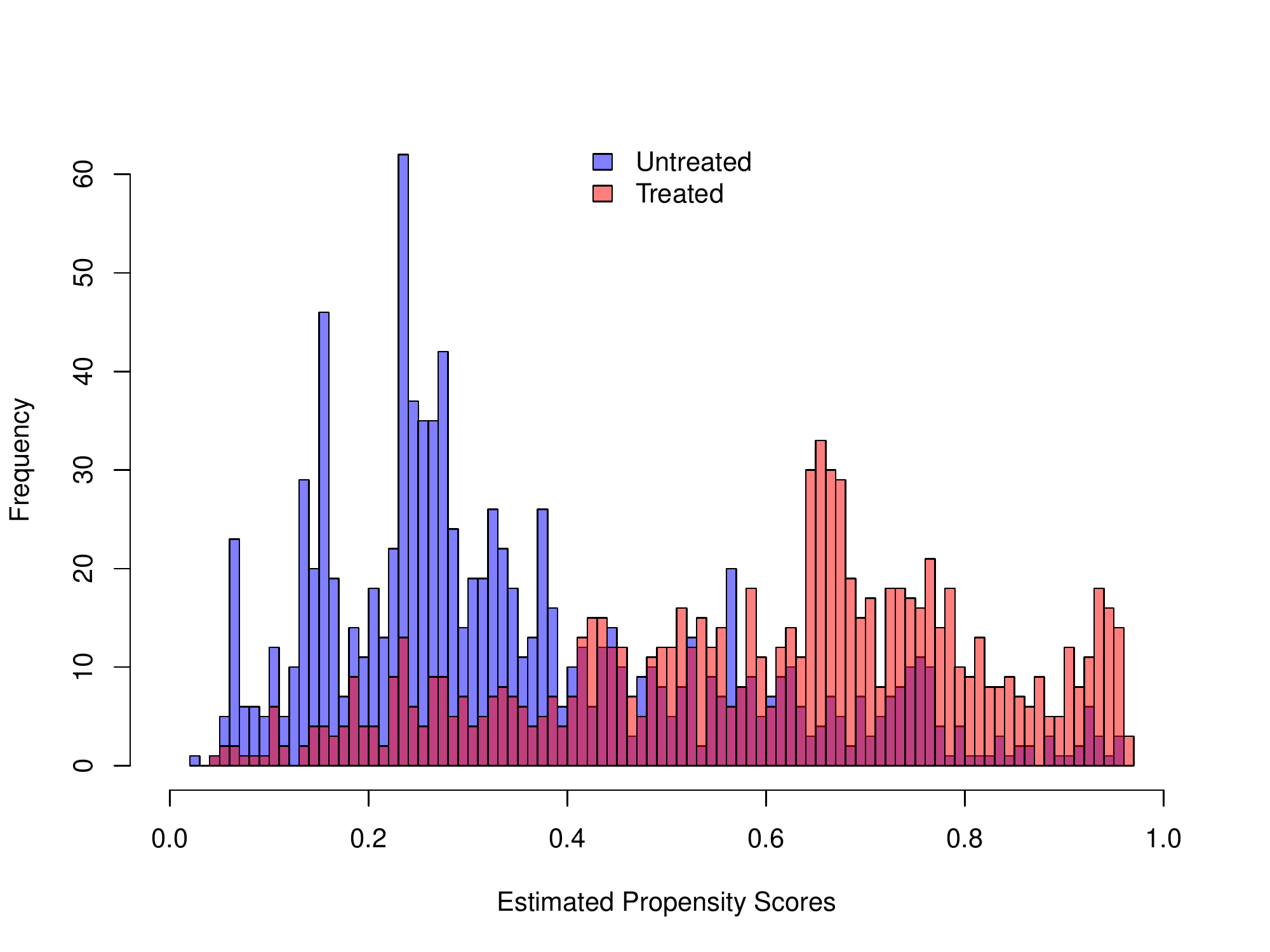}\label{fig:ps2014}}\\
\caption{Illustration of propensity score overlap between power plants ``treated'' with SCR/SNCR technology and those ``untreated'' in four different years.}
\label{fig:ps_overlaps}
\end{figure}

We estimate the propensity score predicting the probability of having SCR/SNCR installed in a given year ($A_i=1$) as a function of power plant characteristics during that year ($\bX_i$). For the years 2002 - 2014 we fit separately to data for each year a logistic regression model of the form:
\begin{align*}
	\logit P(A_i=1 | X_{i}) &= \gamma_0 + \bX_i \gamma,
\end{align*}
where $\bX_i$ includes measures of: annual operating time, annual heat input, percent operating capacity, an indicator of whether the unit participated in Phase II of the Acid Rain Program, the average number of other NO$_x$ controls installed (a single EGU can install multiple technologies), whether the unit is coal-fired with a scrubber, and whether the unit is coal-fired without a scrubber.  Figure \ref{fig:ps_overlaps} depicts the overlap of propensity score distributions for four different years, illustrating the limited overlap in earlier years relative to later years.

\subsection{Balance Checking in the Evaluation of NO$_x$ Emissions-Reduction Technology}
\begin{figure}[htbp]
    \centering
    \includegraphics[width=\linewidth]{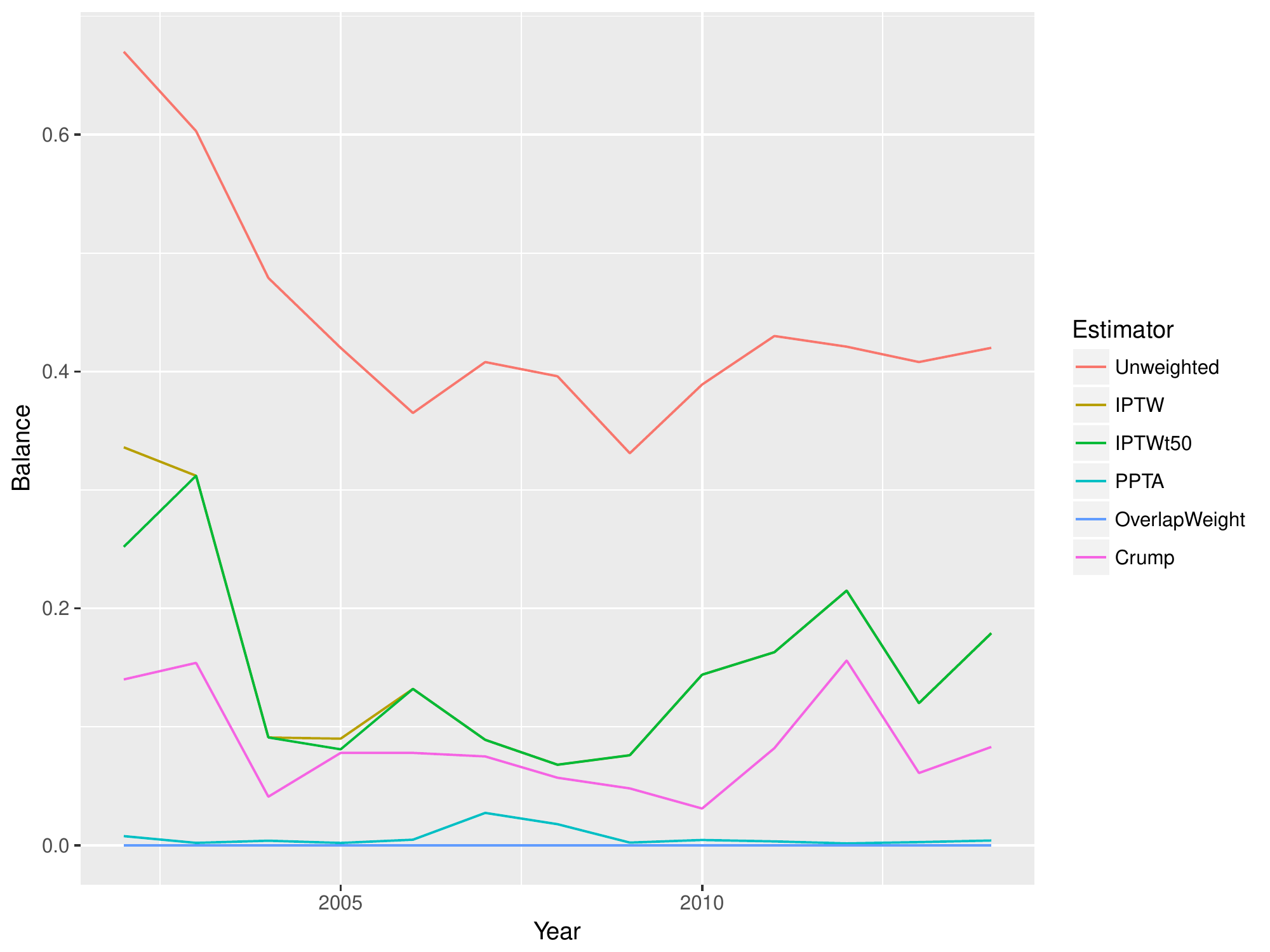}
    \caption{Maximum absolute standardized difference across all covariates by year.}
    \label{fig:mean_bal}
 \end{figure}

Figure \ref{fig:mean_bal} shows the maximum absolute standardized difference in the covariate distribution for each method by year. Note that the PPTA approach averages the standardized difference across all MCMC iterations for each covariate before taking the maximum of the absolute values across covariates. All methods improve balance in comparison to the unweighted analysis. IPTW (with or without truncation) struggles to balance all of the covariates, with at least one covariate having a standardized difference above 0.2 in three years (2002, 2003, and 2012). The method of \cite{crump_dealing_2009} performs well, with maximum absolute standardized differences of less than 0.2 in all years.  PPTA approach achieves standardized differences very close to zero across all years and all covariates, showing only slight differences with the overlap weights that perfectly balance the covariates (with a fixed propensity score) by design. 

\begin{figure}
\includegraphics[width = \textwidth]{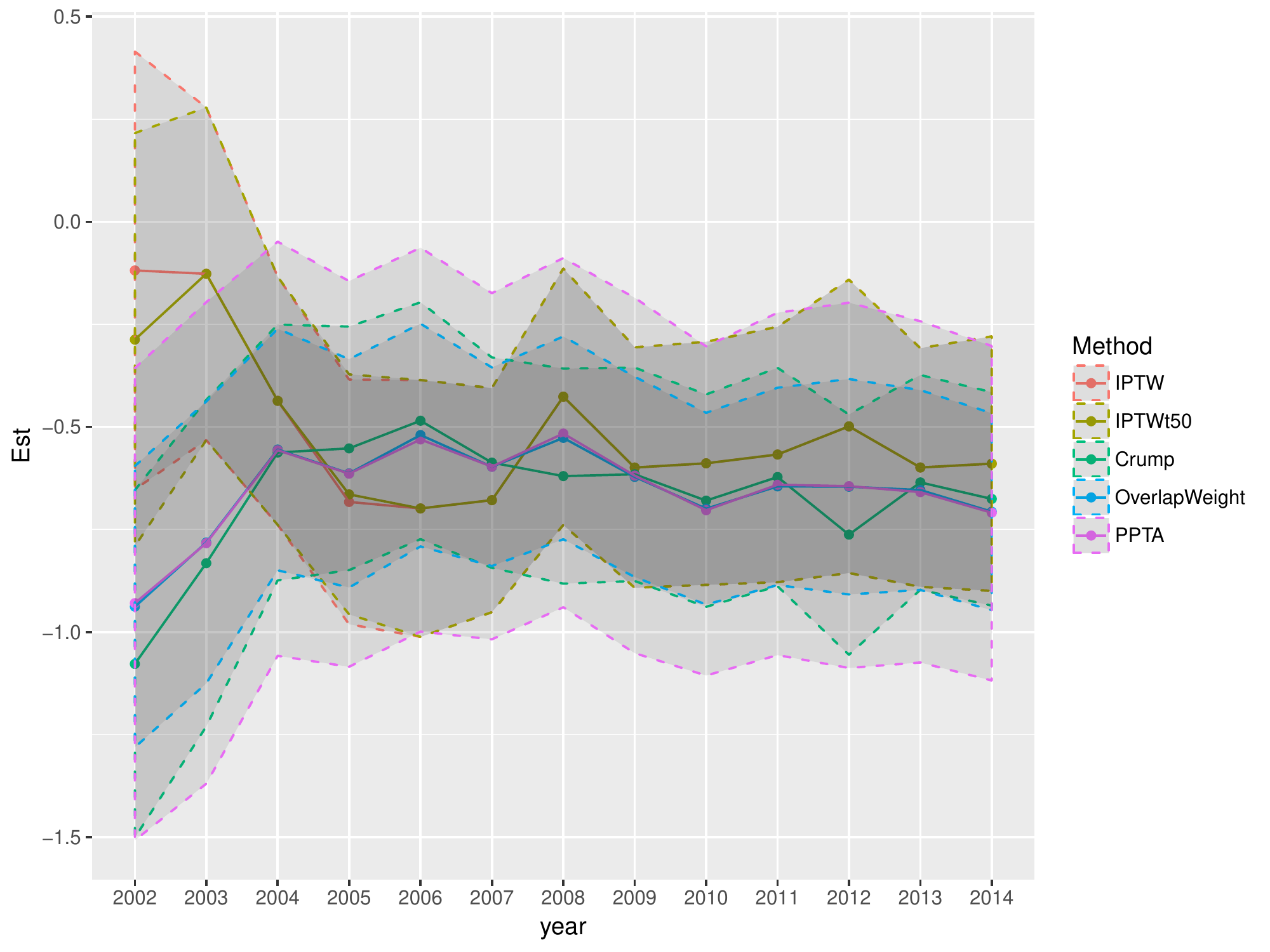}
\caption{Point estimates and 95\% uncertainty intervals}\label{fig:results}
\end{figure}

Figure \ref{fig:results} depicts point estimates and 95\% uncertainty intervals for each estimate across all years.  Point estimates from the PPTA approach are posterior means, and uncertainty intervals are the (2.5\%, 97.5\%) posterior quantiles.  In 2002 and 2003, when overlap is limited, IPTW and IPTW50 estimate no effect of SCR/SNCR on NO$_x$ emissions, a result that is not credible given what is known about the technology.  In contrast, all three other methods similarly estimate a significant reduction in NO$_x$ emissions.  In later years, as the problem of limited overlap dissipates, all five methods begin to produce very similar results, all suggesting that SCR/SNCR technology significantly reduces NO$_x$ emissions by approximately the same amount. The more pronounced effect estimated by PPTA, Crump, and overlap weights in 2002 and 2003 likely results from a relatively weaker regulatory environment that prompted only the plants which stood to benefit the most to install SCR/SNCR, but as the regulations strengthened and installation became more incentivized for the entire population, the effect of SCR/SNCR is estimated to be essentially constant over time.  Note that the PPTA approach has systematically larger uncertainty intervals due to its acknowledgment of propensity score estimation uncertainty and subsequent decision about which observations to include for inference.

\section{Discussion}
Strong confounding presents inherent challenges to estimation of causal effects.  When manifest as limited overlap in propensity score distributions, many existing estimators suffer in finite samples.  We have provided a principled approach that can be viewed as a stochastic generalization of existing ad-hoc approaches to prune or pre-process a sample with weight truncation or deletion.  The proposed two-step Bayesian estimation approach adheres to notions of separating design and analysis while propagating uncertainty in design decisions into estimates of causal effects.  The result is posterior estimates of causal effects where the contribution of each observational unit is directly related to the empirical basis for making a causal inference for that unit, rendering an interpretation analogous to that produced by the ``overlap weights'' of \cite{li_balancing_2016}.  In addition to exhibiting favorable performance in a simple simulation study, the proposed PPTA approach yielded results in an analysis of power plant emissions controls that were more credible than those produced by IPTW (with or without weight truncation) in light of what is known about the technology being evaluated.

One important consideration with the proposed PPTA approach pertains to the interpretation of causal quantities marginalized over different subsets of the observed sample.  Strictly speaking, there is no guarantee that the procedure averages over a single quantity; the causal estimand may vary with $\Ssc$.  This can be regarded as a limitation when interest lies in an estimand like the ATE, but the method may still prove valuable when confronted with the practical reality of erratic performance of estimators that are asymptotically unbiased for the ATE. Anchoring the definition of $\Ssc$ to a quantity that relates directly to the extent of overlap for each observation offers a possible interpretation advantage over existing approaches that condition inference on a single subset of the data defined for mathematical convenience (e.g., as with weight truncation of \cite{crump_dealing_2009}).  Rather than confine inference to a single subset of the data, the PPTA approach averages inference over observations in accordance with the likelihood that each observation could have received the opposite treatment, rendering an estimate interpretable as the ATO, proposed by \cite{li_balancing_2016} with additional motivation and mathematical underpinning.  The ATO may be beneficial as an alternative to the ATE or may be of intrinsic interest.

The proposed method uses the propensity score and PPTA to stochastically create subsets of the data for which treatment assignment is not confounded.  The approach is deliberately framed within the rubric a ``design stage'' approximating that of a randomized experiment, as formulated by \cite{rubin_for_2008}.  While the design decisions considered here were restricted solely to the decision of whether to include an observation in a subset, this represents a simplistic characterization of design meant only to focus on issues of overlap.  More generally, many such design decisions can be made towards approximation of a randomized experiment.  The framework for two-stage Bayesian estimation outlined here to marginalize inference for causal effects over uncertainty in design decisions could extend to other types of design decisions.  This is an important area of future work, with preliminary progress in \cite{alvarez_uncertain_2014} who marginalize over different propensity score matches and \cite{zigler_uncertainty_2014} who marginalize over uncertainty in which variables to include in the propensity score.  The salient theme is that, in contrast to existing approaches that condition all inference for causal effects in the analysis stage on a single fixed design (denoted here with $\Ssc$), marginalizing inference for causal effects over the uncertainty in $\Ssc$ acknowledges uncertainty the various choices made towards construction of that design that are not necessarily characterized by standard estimation uncertainty.


\bibliographystyle{biom}
\bibliography{BayesianPS}

\section{Acknowledgements}
This work was supported by funding from NIH R01ES026217 and R01GM111339, HEI 4909 and 4953, and EPA 835872.  Its contents are solely the responsibility of the grantee and do not necessarily represent the official views of the USEPA. Further, USEPA does not endorse the purchase of any commercial products or services mentioned in the publication.

Information regarding data and code to implement the analysis of Section \ref{sec:application} is available at https://osf.io/s98qm/.
 

\pagebreak
 

\pagebreak

\pagebreak
 

\end{document}

%% file: GrandMacros.tex
\def\ba{{\mbox{\boldmath$a$}}}

\def\bx{{\bf x}}
\def\by{{\bf y}}

\def\bS{{\bf S}}

\def\bU{{\bf U}}

\def\bX{{\bf X}}

\def\thick#1{\hbox{\rlap{$#1$}\kern0.25pt\rlap{$#1$}\kern0.25pt$#1$}}

\def\bgamma{\boldsymbol{\gamma}}

\def\smbalpha{\boldsymbol{{\scriptstyle{\alpha}}}}

\def\ahat{{\widehat a}}

\def\Atilde{{\widetilde A}}

\def\smbalpha{\widehat{\smbalpha}}

\def\hbar{\bar{ h}}

\def\Lsc{{\cal L}}

\def\Nsc{{\cal N}}

\def\Ssc{{\cal S}}

\def\E{\mbox{E}}

\def\logit{\mbox{logit}}

\def\mybox#1{\vskip1mm \begin{center}
        \hspace{.0\textwidth}\vbox{\hrule\hbox{\vrule\kern6pt
\parbox{.9\textwidth}{\kern6pt#1\vskip6pt}\kern6pt\vrule}\hrule}
        \end{center} \vskip-5mm}
\def\lboxit#1{\vbox{\hrule\hbox{\vrule\kern6pt
      \vbox{\kern6pt#1\vskip6pt}\kern6pt\vrule}\hrule}}

\def\thickboxit#1{\vbox{{\hrule height 1mm}\hbox{{\vrule width 1mm}\kern6pt
          \vbox{\kern6pt#1\kern6pt}\kern6pt{\vrule width 1mm}}
               {\hrule height 1mm}}}

\def\fat#1{\hbox{\rlap{$#1$}\kern0.25pt\rlap{$#1$}\kern0.25pt$#1$}}